\documentstyle[12pt]{article}
\begin{document}
\def	\be		{\begin{equation}}
\def	\ee		{\end{equation}}
\def	\ba		{\begin{eqnarray}}
\def	\ea		{\end{eqnarray}}
\def	\nn		{\nonumber}
\def	\=		{\;=\;}
\def	\frac		#1#2{{#1 \over #2}}
\def	\ret		{\\[\eqskip]}
\def	\to		{\rightarrow }
\def	\ie		{{\em i.e.\/} }
\def	\eg		{{\em e.g.\/} }
\def	\g		{\mbox{$\gamma$}}
\def	\e		{\mbox{$e$}}
\def	\m		{\mbox{$\mu$}}
\def	\b		{\mbox{$b$}}
\def	\bbar		{\mbox{$\bar b$}}
\def	\jpsi		{\mbox{$\psi$}}
\def	\psip		{\mbox{$\psi^\prime$}}
\def	\ptmin		{$p_T^{min}$}
\def	\pt		{\mbox{$p_T$}}
\def	\Pt		{\mbox{$P_T$}}
\def	\et		{\mbox{$E_T$}}
\def	\as		{\mbox{$\alpha_s$}}
\def	\mur		{\mbox{$\mu_{\rm{R}}$}}
\def	\muf		{\mbox{$\mu_{\rm{F}}$}}
\def	\mufr		{\mbox{$\mu_{\rm{frag}}$}}
\def	\muzero		{\mbox{$\mu_0$}}
\pagestyle{empty}
\rightline{\vbox{
\halign{&#\hfil\cr
&hep-ph/9405407 \cr
&FERMILAB-PUB-94/135-T\cr
&NUHEP-TH-94-11\cr
&OCIP/C-94-3\cr
& May 1994\cr}}}
\bigskip
\bigskip
\bigskip
{\Large\bf
	\centerline{Fragmentation production of}
        \centerline{$J/ \psi$ and $\psi'$ at the Tevatron}}
\bigskip
\normalsize
\centerline{Eric Braaten\footnote{on leave from the Dept. of Physics and
Astronomy, Northwestern University, Evanston, IL 60208}}
\centerline{\sl Fermi National Accelerator Laboratory, P.O. Box 500, Batavia,
IL 60510}
\bigskip
\centerline{Michael A. Doncheski}
\centerline{\sl Department of Physics,
                Carleton University, Ottawa, Ontario K1S 5B6, Canada}
\bigskip
\centerline{Sean Fleming}
\centerline{\sl Department of Physics and Astronomy,
                Northwestern University, Evanston, IL 60208}
\bigskip
\centerline{Michelangelo L. Mangano}

\centerline{\sl INFN, Scuola Normale Superiore and Dipartimento di Fisica,
                Pisa, Italy}

\bigskip
\begin{abstract}

We present a calculation of the charm and gluon fragmentation contributions
to inclusive $J/\psi$ and $\psi'$ production at large transverse momentum
at the  Tevatron.  For $\psi$ production, we include both fragmentation
directly into  $\psi$ and fragmentation into $\chi_c$ followed by the
radiative decay $\chi_c \to \psi + \gamma$.  We find that fragmentation
overwhelms the leading-order mechanisms for prompt $\psi$ production at
large $p_T$, and that the dominant contributions come from fragmentation
into $\chi_c$.  Our results are consistent with recent data on $\psi$
production from the CDF and D0 experiments.  In the case of prompt $\psi'$
production, the dominant mechanism at large $p_T$ is charm fragmentation
into $\psi'$.  We find serious disagreement between our theoretical
predictions and recent $\psi'$ data from the Tevatron.

\end{abstract}
\vfill\eject\pagestyle{plain}\setcounter{page}{1}
\section{Introduction}
The study of charmonium production in high energy hadronic collisions
provides an important testing ground for perturbative quantum chromodynamics
(QCD).  The $J/\psi$ and $\psi'$ states are of particular interest since
they are produced in abundance and are relatively easy to detect at a
collider such as the Tevatron.  In earlier calculations of direct charmonium
production at large transverse momentum ($p_T$) in $p \bar{p}$ collisions
\cite{br}\ , it was assumed that the leading-order diagrams give the dominant
contributions to the cross section.  These calculations did not reproduce all
aspects of the available data \cite{ua1,cdf88,mlm}, suggesting that there are
other important production mechanisms.  It was pointed out by Braaten and
Yuan \cite{by1} in 1993 that fragmentation processes, while formally of
higher order in the strong coupling constant $\alpha_{s}$, will dominate at
sufficiently large $p_T$.  Explicit calculations of the contribution to
$\psi$ production at the Tevatron from the fragmentation of gluons and charm
quarks revealed that fragmentation dominates over the leading-order
gluon-gluon fusion mechanism for $p_{T}$ greater than about 6 GeV \cite{dfm}.
In addition to being directly produced, the $\psi$ signal is also fed by the
$\chi_c$ states through the radiative decay $\chi_c\to\psi +\gamma$.  In
this paper, we present more complete calculations for $\psi$ production,
including the effects of $\chi_c$'s that are produced by fragmentation,
and compare the results with recent data on prompt and inclusive $\psi$
production from the CDF and D0 experiments.  We also compare our
calculations of $\psi'$ production with recent data from CDF.

While this paper was being written, similar work on $\psi$-production was
presented in a paper by Cacciari and Greco \cite{cg}.

\section{Fragmentation Formalism}

Factorization theorems of perturbative QCD indicate that the inclusive
production of a hadron at large $p_{T}$ is dominated by fragmentation.
Fragmentation is the
production of a parton with large transverse momentum which subsequently
forms a jet containing the desired hadron.  In the case of $p \bar{p}$
collisions, the fragmentation contribution to the cross section can be
expressed as a convolution of parton distribution functions, hard-scattering
cross sections, and fragmentation functions.  Taking $\psi$ production to be
specific, the differential cross section can be written as:
\begin{equation}
d\sigma(p\bar{p} \rightarrow \psi (p_{T},y)+X)=
\sum_{i} \int^{1}_{0} \! dz \; d \sigma(p\bar{p} \rightarrow
i(\frac{p_{T}}{z},y)+X, \mufr^2) \; D_{i \rightarrow  \psi}(z, \mufr^2),
\label{dcs}
\end{equation}
where $z$ is the longitudinal momentum fraction of the $\psi$ relative to
parton $i$, $y$ is the rapidity of the $\psi$, and
$D_{i \to \psi}(z, \mufr^2)$ is the fragmentation function.  The dependence
on the fragmentation scale \mufr\ cancels between the two factors only
after inclusion of all orders of the perturbative expansion.  The
differential cross section on the right side of (\ref{dcs}) can in turn be
written as the convolution of parton distributions $f_{j/p}$ and
$f_{k/\bar{p}}$ in the proton and antiproton with hard-scattering
differential cross sections $d{\hat \sigma}$ for the parton subprocesses
$j + k \to i + X$.  In low-order calculations, the fragmentation scale
\mufr\ , the factorization scale \muf\ , and the renormalization scale \mur\
should all be chosen on the order of $p_T/z$, the transverse momentum of the
fragmenting parton.  The dominant contributions to Eq. (\ref{dcs}) come from
gluon fragmentation and charm fragmentation.  Fragmentation of light quarks
will only contribute at low $z$, and is therefore strongly suppressed by the
rapidly falling \pt\ spectrum of the final-state partons from the
hard-scattering process.  The inclusive $\psi$ signal is also fed by
production of the P-wave $\chi_c$ states followed by the radiative decay
$\chi_c \to \psi + \gamma$.  The fragmentation contributions can be obtained
by multiplying the $\chi_c$ production cross sections analogous to
(\ref{dcs}) by the radiative branching fractions 0.007, 0.27, and 0.135 for
$\chi_{c0}$, $\chi_{c1}$, and $\chi_{c2}$, respectively.

The fragmentation functions $D(z)$ for charmonium production can be
calculated within perturbative QCD \cite{by1,cc}.  The relevant fragmentation
functions for the production of the S-wave and P-wave states have all been
calculated to leading order in $\alpha_s$.  In this paper we use the
fragmentation functions for $g \to \psi$ \cite{by1}, $c \to \psi$ \cite{bcy},
$g \to \chi_c$ \cite{by2}, $c \to \chi_c$ \cite{y}, and $\gamma \to \psi$
\cite{sf}.  The perturbative calculations give the fragmentation functions at
an initial scale $\mu_0$ of order $m_c$.  We take this initial scale to be
$\mu_0=2m_c$ for gluon and photon fragmentation, and $\mu_0=3m_c$ for charm
fragmentation.  The fragmentation functions are then evolved up to the
scale $\mufr = {\cal O} (p_T/z)$ set by the transverse momentum of the
fragmenting  parton using the Altarelli-Parisi evolution equations:
\begin{equation}
\mu^2 \frac{\partial}{\partial \mu^2} D_{i \rightarrow \psi}(z, \mu^2) =
\frac{\alpha_s}{2\pi}
\sum_{j} \int^{1}_{z} \frac{dy}{y} \; P_{ij}(z/y)
\; D_{j \rightarrow \psi }(y, \mu^2).
\label{ap}
\end{equation}
We included in Eq. (\ref{ap}) only the $P_{gg}$ splitting term for gluon
fragmentation and only the $P_{cc}$ term for charm fragmentation.  With the
exception of $D_{g\to\psi}$, the inclusion of the off-diagonal
Altarelli-Parisi kernels would give corrections that are safely negligible
within the accuracy of a leading-order (LO) calculation.  In the case of
$D_{g\to \psi}$, the $P_{gc}$ term in (\ref{ap}) is important because the
initial fragmentation function $D_{c\to\psi}(z,\mu_0^2)$ is more than an
order of magnitude larger than $D_{g\to\psi}(z,\mu_0^2)$ \cite{wise}.  In
the present analysis, we take this effect into account by including the
next-to-leading-order (NLO) corrections to the hard-scattering cross
sections $d {\hat \sigma}$ for inclusive charm production \cite{NDE}.  These
higher order terms include the numerically important contributions from
final state gluon splitting into $c\bar c$ pairs.  Having included this
effect in the hard-scattering cross sections, we do not have to account for
them via the $g\to c\bar c$ Altarelli-Parisi kernel as in Ref. \cite{cg}.
Higher-order terms from off-diagonal Altarelli-Parisi evolution, such as
$g \to c \bar c$ splitting from secondary gluons in the gluon-jet cascade,
contribute to the overall charmonium multiplicity inside the jet, but the
enhancement comes only in the small $z$ region and it can therefore be safely
neglected in the LO \pt\ distribution.

We now discuss the uncertainties in the calculations of the fragmentation
contribution to the differential cross section for charmonium production.
We first consider the initial fragmentation functions $D(z,\mu_0^2)$.  They
have been calculated only to leading order in $\alpha_s$ and to leading
order in a nonrelativistic expansion.  Based on the NLO calculations of the
annihilation rates for charmonium, we anticipate that NLO corrections to the
fragmentation functions may be as large as 50\%.  A rough estimate of the
size of relativistic corrections is the average value of $v^2/c^2$ in
potential models, which is about 30\%.  The normalizations of the initial
fragmentation functions are calculated in terms of $\alpha_s$, $m_c$, and
various nonperturbative matrix elements.  For the charm quark mass, we use
the value $m_c = 1.5$ GeV.  While $m_c$ appears in the fragmentation
functions raised to the third or fifth power, this is not a large source of
uncertainty since roughly the same power appears in the quantities that are
used to determine the nonperturbative matrix elements.  The S-wave
fragmentation functions depend on the wavefunction at the origin, which we
take to be $|R_{\psi}(0)|^2 = 0.7 \; {\rm GeV}^3$.  This value is obtained
from the electronic width of the $\psi$, including the effect of the NLO
perturbative correction, which is about 50\%, but not taking into account
relativistic corrections.  The P-wave fragmentation functions depend on
two nonperturbative parameters.  For the derivative of the radial
wavefunction at the origin, we use the value
$|R'_{\chi_c}(0)|^2 = 0.053 \; {\rm GeV}^5$.  This value is determined from
the annihilation rates of the $\chi_c$ states \cite{bbl}, neglecting the
as-yet-uncalculated NLO perturbative corrections as well as relativistic
corrections.  The least well-determined parameter in the P-wave
fragmentation functions is a parameter $H'_8$ associated with the
color-octet mechanism for P-wave production \cite{bbly}.  This parameter
is poorly constrained, lying in the range $2.2 < H'_8 < 25$ MeV
\cite{by2,ht}.  We use the value $H_8' = 3 \; {\rm MeV}$, which is
consistent with measured branching fractions for $B$ mesons into the
$\chi_c$ states \cite{bbly}.

We next consider errors due to the evolution of the fragmentation functions
from the initial scale $\mu_0$ up to the scale $p_T/z$.  The
Altarelli-Parisi equations Eq. (\ref{ap}) break down in the small-$z$ region,
due to large logarithms of $1/z$ in the perturbation expansion.  The most
dramatic effect of this breakdown is an unphysical divergence in the gluon
multiplicity  at low momentum fraction \cite{bcm}.  This leads to a
correspondingly large overestimate of the gluon fragmentation functions into
charmonium at small values of $z$.  We do not expect this to be a serious
problem at the values of $p_T$ considered in this paper, because the cross
sections are dominated by larger values of $z$ due to the steeply falling
spectrum of the hard partons.  We find empirically that the Monte Carlo
calculations presented below rarely sampled values of $z$ smaller than 0.1.

While the divergence of the gluon multiplicity at small $z$ may not in itself
be a problem, it is a symptom of a deficiency of Altarelli-Parisi evolution
that also has consequences at larger values of $z$.  In particular, the
naive Altarelli-Parisi equations  do not respect the phase-space constraint
$D_{g \to \psi}(z,\mu^2)=0$ for $z < M_\psi^2/\mu^2$.  The implementation of
this  constraint would slow down the evolution of the fragmentation function
by delaying the depletion of the large-$z$ fragmentation region.  Since the
spectrum of gluons and charm quarks falls rapidly with $p_T/z$, a proper
treatment of the large-$z$ region can have a significant effect on the cross
section.  A more accurate treatment would be to use the following system of
equations for the evolution of the gluon fragmentation functions:
\ba        \label{ap+}
D(z,\mu^2) &=& \int_{M_\psi^2}^{\mu^2} \; \frac{dq^2}{q^2}
\int_{z}^1 \frac{dy}{y} G(y,q^2;\mu^2) d(z/y,q^2)  \\
\label{ap++}
\mu^2 \frac{\partial \ }{\partial \mu^2} G(z,\mu^2;\mu_0^2) &=&
\frac{\alpha_s}{2\pi}
\int^{1}_{z} \frac{dy}{y} \; P_{gg}(y)
\; G(z/y,\mu^2; y \mu_0^2)
\ea
where $G(z,\mu^2;\mu_0^2)$ is the distribution of gluons of virtuality $\mu$
inside a gluon of virtuality $\mu_0$, which is subject to the boundary
condition $G(z,\mu_0^2;\mu_0^2) = \delta(1-z)$.  The function
$d(z,q^2)$~\cite{by1} is
the LO probability that a gluon of virtuality $q$ decays to a \jpsi\ carrying
longitudinal momentum fraction $z$ in the infinite momentum frame.  The
initial fragmentation functions in our present treatment are given by
$D(z,\mu_0^2)= \int_0^\infty (ds/s) d(z,s)$.  The system of equations
(\ref{ap+}) and (\ref{ap++}) gives the correct $\psi$ multiplicity inside a
gluon jet \cite{mlm2}, after inclusion of small-$z$ coherence effects
\cite{bcm}.
It is straightforward to show that this
system is equivalent to the following nonhomogeneous evolution equation:
\be \label{nhap}
\mu^2 \frac{\partial}{\partial \mu^2} D(z,\mu^2) \;=\;
d(z,\mu^2) \;+\; \frac{\alpha_s}{2\pi}
\int^{1}_{z} \frac{dy}{y} \; P_{gg}(y)
\; D(z/y, y \mu^2)  ,
\ee
together with the boundary condition $D(y,\mu^2=M_{\psi}^2) = 0$.  This
evolution equation respects the phase space constraint, as can be easily
checked \cite{bcm}.  A thorough study of this generalized evolution equation
and its consequences for charmonium production will be presented elsewhere.

The evolution equation presented above also solves a problem involving
threshold effects in our fragmentation functions.  The \pt\ values for which
the fragmentation contributions become important are, in fact, too small to
be considered in the asymptotic regime where threshold effects can be
ignored.  In the calculations of the initial fragmentation functions
$D(z,\mu_0^2)$, the assumption $\mu_0^2 \gg 4m_c^2/z$ was used to obtain
simple analytic results.  This assumption is not really compatible with the
subsequent identification $\mu_0 = 2 m_c$.  The resulting
error may be negligible after evolution to asymptotically large $\mu$, but it
is probably significant at the scales needed in our calculations.  The
evolution equation (\ref{nhap}) treats threshold effects consistently.  In
the present analysis, we estimate the error due to our treatment of
threshold effects by determining the change in the cross section that results
from  increasing the value of $\mu_0$ by a factor of 2.

There are several other sources of uncertainty.  The uncertainties due to the
choice of the renormalization scale, the factorization scale, and the
fragmentation scale can be estimated by varying these scales by factors of 2.
The error from the parton distributions can be estimated by repeating the
calculations using different parton distributions.  Another source of error at
small $p_T$ is the neglect of the intrinsic transverse momentum of the partons
in the proton and antiproton \cite{ce,32}. The effect of the intrinsic
transverse momentum is most significant for partons with very small
longitudinal momentum fraction $x$. It should not be important at large $p_T$
because $\psi$ production at large $p_T$ does not probe deeply into the
small-$x$ region of the parton distribution. Finally there is the NLO
perturbative correction to the hard-scattering cross section
$d {\hat \sigma}$, which we only included in the case of charm fragmentation.
The NLO corrections to the gluon \pt\ spectrum have been evaluated in
ref.~\cite{cg}\ , where it was shown that they increase the rate for $\chi$
production from fragmentation by about 50\%.  Needless to say, a full NLO
calculation of both the hard-scattering cross sections and the fragmentation
functions would be highly nontrivial.

Considering all the uncertainties discussed above, we believe that the error
in our fragmentation calculations can easily be larger than a factor of 2,
but it is definitely less than an order of magnitude.  One should of course
keep in mind that in addition to fragmentation, which must dominate at
sufficiently large $p_T$, there are other contributions suppressed by
factors of $m^2_c/p^2_T$ that may be important at the values of $p_T$ that
are available experimentally.

\section{Results and discussion}

In figure 1 we plot the individual contributions to the differential cross
section for prompt $\psi$ production as a function of $p_T$.  We include
results for both the fragmentation contributions and the leading-order
contributions\footnote{The results for direct \jpsi\ inclusive \pt\
distributions given in ref.~\cite{mlm} are incorrect, due to a coding error
in the choice of \mur\ and \muf.}.  Note that each of the $\chi_c$
production curves is a sum over the three P-wave states $\chi_{c0}$,
$\chi_{c1}$, and $\chi_{c2}$.  We used the MRSD0 parton  distribution set,
and chose the renormalization (\mur), factorization (\muf) and fragmentation
(\mufr) scales to be same, and equal to the transverse momentum of the
fragmenting parton, \Pt=\pt/$z$.  In order to compare with available data
from the Tevatron, we imposed a pseudorapidity cut of $| \eta | < 0.6$ on
the $\psi$.  It is evident from the graph that fragmentation dominates over
the leading-order mechanisms for all values of $p_T$ for which the
fragmentation approximation is reasonable, namely for $p_T$ greater than
about 5 GeV.  The dominant production mechanism by an order of magnitude is
gluon fragmentation into $\chi_{c}$ followed by its decay into $\psi$.  Note
that, aside from photon fragmentation which is dominated by
quark-gluon initial states, all the fragmentation contributions have
the same $p_T$-dependence.
The leading-order $\chi_c$ production falls off more rapidly with $p_T$,
and that leading-order $\psi$ production falls off still more rapidly.  This
pattern simply reflects the $p_T$-dependence of the underlying
hard-scattering processes: $d {\hat \sigma}/dp_T^2$ scales like $1/p_T^4$ for
fragmentation, $1/p_T^6$ for leading-order $\chi_c$ production, and
$1/p_T^8$ for leading-order $\psi$ production.

In figure 2, the sum of the fragmentation contributions (two solid curves)
and the sum of the leading-order contributions (two dashed curves) are
compared with  preliminary CDF data for prompt $\psi$ production \cite{vaia}.
The contribution to \jpsi\ production from $b$-hadron decays has been
removed from the data via detection of the secondary vertex from which the
\jpsi's originate \cite{vaia}. The upper and lower curves in figure~3 were
obtained by varying the scales \mur, \muf\ and \mufr\ used in the
calculation, in order to provide an estimate of the systematic uncertainty
associated with the LO calculation.  The upper curve corresponds to
\mur=\muf=\Pt/2 and \mufr=max(\Pt/2,\muzero), while the lower curve is
obtained for \mur=\muf=\mufr=2\Pt.  The cross-over of the curves at small $p_T$
is due to the rapid growth of the parton distribution functions with
increasing scale, and should be considered an artificial reduction of scale
sensitivity. The errors from varying the parton
distributions are only about 10\%.  We estimate the error from threshold
effects in the fragmentation functions by increasing $\mu_0$ by a factor
of 2. The effect of this is a decrease of the fragmentation contribution, by a
factor of 2, at $p_T=3$ GeV, and an increase, by a factor of 2, at
$p_T=20$ GeV. Another large uncertainty comes from the color-octet
parameter $H_8'$ in the fragmentation functions for $g \to \chi_c$.
Changing $H'_8$ by a factor of two  changes our results by a factor of 2 at
the largest $\pt$ available.  While the shapes of the leading-order curve
and the fragmentation curve are both consistent with the data over the range
of $p_T$ that is available, the normalization of the leading-order
contribution is too small by more than an order of magnitude.  The
fragmentation contribution has the correct normalization to within a factor
of 2 or 3, which can be easily accounted for by the uncertainties discussed
above.  We conclude that the fragmentation calculation is not inconsistent
with the CDF data on prompt $\psi$ production.

We also present in fig. 3 a comparison of the theoretical predictions for
fully inclusive \jpsi\ production (including those from $b$-hadron decays)
with data from CDF \cite{vaia}\ and D0 \cite{bazizi}.  The theoretical
contribution from $b$-decays used here was evaluated at NLO as in Ref.
\cite{mlm}.  The predictions are consistent with the data.

We next consider the production of $\psi'$, which should not receive any
contributions from higher charmonium states.  The $\psi'$ fragmentation
contribution can be obtained from the $g \to \psi$, $c \to \psi$, and
$\gamma \to \psi$ fragmentation contribution simply by multiplying by the
ratio of the electronic widths of the $\psi'$ and $\psi$.  The total
fragmentation contribution (two solid curves) and the leading-order
contribution (two dashed curves) are shown in figure 4, along with the
preliminary CDF data \cite{troy}.  Again the contribution from $b$--hadron
decays has been subtracted using the secondary vertex information.  The
pairs of curves correspond to the same choices of scales as in figure~2.
The dominant production mechanisms are gluon-gluon fusion for $p_{T}$ below
about 5 GeV, and charm quark fragmentation into $\psi'$ for larger $p_T$.
The leading-order curve falls much too rapidly with $p_T$ to explain the
data, but the fragmentation curve has the correct shape.  However, in
striking contrast to the case of $\psi$ production, the normalization of
the fragmentation contribution to $\psi'$ production is too small by more
than an order of magnitude.  That there is such a large discrepancy between
theory and experiment in the case of $\psi'$, but not for $\psi$, is
extremely interesting.  It suggests that there are other important mechanisms
for production of S-wave states at large $p_T$ beyond those that have
presently been calculated.  While such processes would certainly affect
\jpsi\ production as well, their effect may not be as dramatic because of
the large contribution from $\chi_c$-production in the case of the \jpsi.

One possible such mechanism is the process $g g \to \psi g g$, with a gluon
exchanged in the $t$-channel.  This is a subset of the NLO corrections to
the process $gg \to \psi g$ for which the hard-scattering cross section has
a $p_T$-dependence that is intermediate between the leading-order diagrams
and the fragmentation contribution.  It is easy to verify that
$d{\hat \sigma}/dp_T^2$ for this process scales asymptotically like
$\as^4/\pt^6$, compared to $\as^3/\pt^8$ for $gg \to \psi g$ and compared
to $\as^5/\pt^4$ for the NNLO correction, which includes gluon fragmentation.
The shape of the $p_T$ distribution from this process should be similar to that
for leading-order $\chi_c$ production, which is compatible with the data.
Whether the normalization agrees can only be determined by explicit
calculation.  Such a calculation is in progress.

Note that, like the decay of a $b$-hadron, the fragmentation mechanism
produces $\psi$'s inside a jet of light hadrons.  A non-isolation cut on
the $\psi$ can therefore not be used to tag $\psi$'s coming from $B$ meson
decay.  One might hope that an isolation cut could be used to separate
prompt $\psi$'s produced by the leading-order mechanisms from those
produced by fragmentation.  However, we found in our calculations that the
average value of $z$ for a $\psi$ produced by fragmentation is
$\langle z \rangle = 0.7$.  This means that the remaining partons in the
jet containing the $\psi$ share on average less than 1/2  the energy of the
$\psi$, and thus may often be too soft to be detected reliably.

\section{Conclusion}

We have calculated the cross section for production of prompt $\psi$ at large
$p_T$ at the Tevatron including the effects of gluon and charm fragmentation.
The largest contributions by an order of magnitude come from gluon
fragmentation into $\chi_c$, followed by the decay $\chi_c \to \psi + \gamma$.
The results of this calculation are not inconsistent with preliminary CDF
data, given the uncertainties in the fragmentation functions.  In contrast,
the leading-order mechanisms give a result that is nearly an order of
magnitude too small.

We have also calculated inclusive $\psi'$ production at large $p_T$ at the
Tevatron.  We find that fragmentation dominates for $p_T$ greater than about
6 GeV, with the largest contribution coming from charm fragmentation into
$\psi'$.  Comparing the results of the calculation to preliminary CDF data,
we find that the cross section is too small by more than an order of
magnitude, even after including the fragmentation mechanism.  Thus while the
fragmentation mechanism may provide an explanation for the observed rate of
prompt $\psi$ production at large $p_T$, it does not seem to explain the
existing data for $\psi'$. We discussed the possibility of additional
mechanisms for charmonium production at large $p_T$ beyond those that have
presently been calculated.  Furthermore, we discussed some of the intrinsic
theoretical uncertainties present in these calculations and anticipated some
studies that can be undertaken to improve the present theoretical framework.

The work of M.A.D. is supported in part by the Natural Sciences and
Engineering Research Council of Canada through a Canada International
Fellowship.  The work of M.L.M. is supported in part by the EEC Programme
``Human Capital and Mobility'', Network ``Physics at High Energy Colliders'',
contract CHRX-CT93-0537 (DG 12 COMA).  The work of E.B. and S.F. was
supported in part by the U.S. Department of Energy, Division of High Energy
Physics, under Grant DE-FG02-91-ER40684.  E.B. would like to
thank the Fermilab theory group for their hospitality while this work was
being carried out.

\vfill\eject

\noindent{\Large\bf Figure Captions}
\begin{enumerate}
\item Contributions to the differential
	cross section for inclusive  $\psi$ production at the Tevatron:
	fragmentation into $\psi$ (solid curves),
	and the leading order contributions (dashed curves).
\item Preliminary CDF data for prompt $\psi$ production (O)
	compared with theoretical predictions of
	the total fragmentation contribution (solid curves)
	and the total leading-order contribution (dashed curves).
\item Total $\psi$ production: CDF (O) and D0 ($\Diamond$) data
        compared to theoretical curves for prompt $\psi$ production (solid
        curves), and theoretical predictions for $b$-hadron decays (dashed
        curves).
\item Preliminary CDF data for prompt $\psi'$ production (O)
	compared with theoretical predictions of
	the total fragmentation contribution (solid curves)
	and the total leading-order contribution (dashed curves).
\end{enumerate}
\vfill\eject

\end{document}